# Raman spectroscopy and imaging of graphene

Zhen hua Ni, Ying ying Wang, Ting Yu, and Ze xiang Shen[*]

*Division of Physics and Applied Physics, School of Physical and Mathematical Sciences, Nanyang Technological University, Singapore 637371, Singapore*

**Abstract**

Graphene has many unique properties that make it an ideal material for fundamental studies as well as for potential applications. Here we review the recent results on the Raman spectroscopy and imaging of graphene. Raman spectroscopy and imaging can be used as a quick and unambiguous method to determine the number of graphene layers. Following, the strong Raman signal of single layer graphene compared to graphite is explained by an interference enhancement model. We have also studied the effect of substrates, the top layer deposition, the annealing process, as well as folding (stacking order) on the physical and electronic properties of graphene. Finally, Raman spectroscopy of epitaxial graphene grown on SiC substrate is presented and strong compressive strain on epitaxial graphene is observed. The results presented here are closely related to the application of graphene on nano-electronic device and help on the better understanding of physical and electronic properties of graphene.

*Corresponding author: zexiang@ntu.edu.sg

## 1. Introduction

Graphene comprises one monolayer of carbon atoms packed into a two-dimensional (2D) honeycomb lattice.[1] It has attracted much interest since it was firstly discovered in 2004.[2] Graphene is the basic building block for other carbon nanomaterials, such as 0D fullerenes, 1D carbon nanotubes and 2D nanographite sheets. In the electronic band structure of graphene, the conduction band touches the valence band at two points (*K* and *K'*) in Brillouin zone, and in the vicinity of these points, the electron energy has a linear relationship with the wavevector, $E = \hbar k v_f$. Therefore, electrons in an ideal graphene sheet behave like massless Dirac-Fermions.[1] The unique properties of graphene make it a promising candidate for fundamental study as well as for potential device applications.[3-8] The charge carriers in graphene can be tuned continuously between electrons and holes in concentrations *n* as high as $10^{13}$ cm$^{-2}$. The mobility of graphene at room temperature is ~ 120, 000 cm$^2$/Vs, higher than in any known semiconductor.[9, 10] Besides, the application of graphene on spintronics has also received a lot of interest.[11, 12] The spin relaxation length of ~2 um has been observed in graphene, which gives a promising future of gaphene spintronics. Many other applications of graphene have also been reported.[5, 12-15]

Raman spectroscopy has been historically used to probe structural and electronic characteristics of graphite materials, providing useful information on the defects (D band), in-plane vibration of sp$^2$ carbon atoms (G band) as well as the stacking orders (2D band).[16-18] Raman spectroscopic studies of graphene have revealed very interesting phenomena. For example, the single and sharp second order Raman band (2D) has been widely used as a simple and efficient way to determine the single layer graphene (SLG).[19] The 2D band of multilayer graphene can be fitted with multiple peaks, which is due to the splitting of electronic band structure of multi-layer graphene.[20] The electronic structure of bilayer graphene (BLG) is also probed by resonant Raman scattering.[21] The electrical field effect studies have revealed that electron/hole doping in graphene will affect the electron-phonon coupling, hence the Raman frequency.[22-24] The in-plane vibrational G band of graphene blueshifts in both electron and hole doping, due to the nonadiabatic removal of the Kohn anomaly at Γ point. On the other hand, the 2D band redshifts (blueshifts) for electron (hole) doping, due to the charge transfer induced modification of the equilibrium lattice parameter.[25] This makes Raman spectroscopy an effective technique to determine the doping type and doping concentration in graphene.[26, 27] The effects of temperature on Raman spectra of graphene have also been studied and the lattice inharmonicity of graphene is investigated.[28] Moreover, the extremely high thermal conductivity of graphene (~3080-5150W/mK) has recently been obtained by Raman spectroscopy, which suggests the potentiall application of graphene as thermal management material in future nanoelectronic devices.[29]

In this contribution, the Raman spectroscopy and imaging are used to study the graphene in several aspects: (i) Raman spectroscopy and imaging can be used as an efficient way to determine the number of graphene layers. (ii) The strong signal of graphene on $SiO_2$/Si substrate is explained by an interference enhancement model. (iii) The effect of substrates, top insulator deposition and annealing process, as well as folding and stacking order on physical and electronic structure of graphene are studied by Raman spectroscopy and imaging. (iv) Raman studies of epitaxial graphene grown on SiC substrate are also presented.

## 2. Preparation of graphene

Several approaches have been successfully developed to fabricate graphene, such as micromechanical cleavage of graphite [2] and epitaxial growth on silicon carbide (SiC) substrate.[30, 31] The former can be used to obtain high quality graphene sheets which are comparable to that in graphite, but is restricted by small sample dimensions and low visibility. Epitaxial graphene (EG) grown on SiC is suitable for large area fabrication and is more compatible with current Si processing techniques for future applications. However, the quality of EG still needs to be improved and the effects of substrate on its properties are not well understood. In addition to the above methods, a number of work also report the growth of graphene on metal substrates, such as ruthenium [32] and nickel,[33] but this would require the sample to be transferred to insulating substrates in order to make useful devices. There are also liquid phase exfoliation approaches to make graphene sheets film by dispersion and exfoliation of graphite in organic solvents.[13, 34] Chemical routes are also developed to fabricate graphene and graphene nanostructures and interesting results are reported.[35, 36] In this paper, the samples are prepared by micromechanical cleavage or epitaxial grown on SiC substrate.

## 3. Raman spectroscopy and imaging of graphene and layer thickness determination

A quick and precise method for determining the number of layers of graphene sheets is essential for speeding up the research and exploration of graphene. Although atomic force microscopy (AFM) measurement is the most direct way to identify the number of layers of graphene, the method has a very slow throughput. Furthermore, an instrumental offset of ~0.5 nm (caused by different interaction force) always exists, which is even larger than the thickness of a monolayer graphene and a data fitting is required to extract the true thickness of graphene sheets.[37] Unconventional Quantum Hall effects[3, 38] are often used to differentiate monolayer and bilayer graphene from few-layers. However, it is not a practical and efficient way. Researchers have attempted to develop more efficient ways to identify different layers of graphene without destroying the crystal lattice. Raman spectroscopy is the potential candidate for non-destructive and quick inspection of number of layers of graphene.[19, 39]

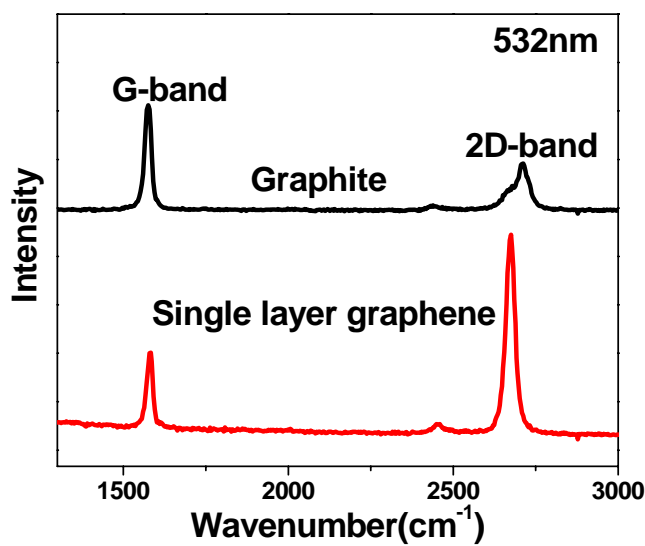

Figure 1. Raman spectra of single layer graphene and graphite.

Figure 1 gives typical Raman spectra of SLG and graphite on a $SiO_2$ (300 nm) /Si substrate. The Raman spectra are recorded with a WITEC CRM200 Raman system with a double-frequency Nd:YAG laser (532 nm) as excitation source. The laser power at sample is below 0.1 mW to avoid laser induced heating. The two spectra are recorded in the same experimental conditions. The Raman signal of SLG is unexpectedly strong and even comparable to that of bulk graphite. This interesting phenomenon will be addressed in section 4 later.[40] The major Raman features of graphene and graphite are the so called G band (~1580 cm$^{-1}$) and 2D band (~2670 cm$^{-1}$). The G band originates from in-plane vibration of sp$^2$ carbon atoms is a doubly degenerate (TO and LO) phonon mode ($E_{2g}$ symmetry) at the Brillouin zone center.[41] The 2D band originates from a two phonon double resonance Raman process.[42] The 2D band is also named as G' band in some work,[20] as it is the second most prominent band of graphite samples in addition to the G band. Here, we refer the band as 2D band because it is the second order overtone of the D band. The obvious difference of the Raman features of SLG and graphite is the 2D band. For SLG, the 2D band can be fitted with a sharp and symmetry peak while that of graphite can be fitted with two peaks. It can be seen in Fig 2 that the 2D band becomes broader and blueshifted when the graphene thickness increases from SLG to multilayer graphene. As the 2D band origins from the two phonon double resonance process, it is closely related to the band structure of graphene layers. Ferrari et al.[19] have successfully used the split of electronic band structure of multi-layer graphene to explain the broadening of 2D band. Cancado et al [20] also gave a detailed discussion on how the multi Raman peak structure of 2D band obtained from stacking of two or more graphene layers is related to the dispersion of π electrons. As a result, the sharp and symmetric 2D band is widely used to identify SLG.[5] Unfortunately, the 2D band differences between two and few layers of graphene sheets are not obvious and unambiguous in the Raman spectra. In addition to the difference of 2D band, the G band intensity increases almost linearly as the graphene thickness increases, as shown in Fig 2. [40] This can be understood as more carbon atoms are detected for multi-layer graphene. Therefore, the intensity of G band can be used to determine the number of layers of graphene reversely.

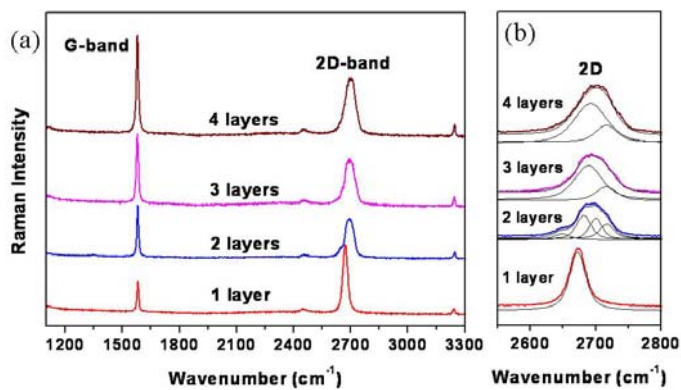

Figure 2. (a) Raman spectra of graphene with 1, 2, 3 and 4 layers. (b) The enlarged 2D band regions with curve fit. Reproduced with permission from *J. Phys. Chem. C* 2008, *112*, 10637-10640. Copyright 2008 American Chemical Society.[43]

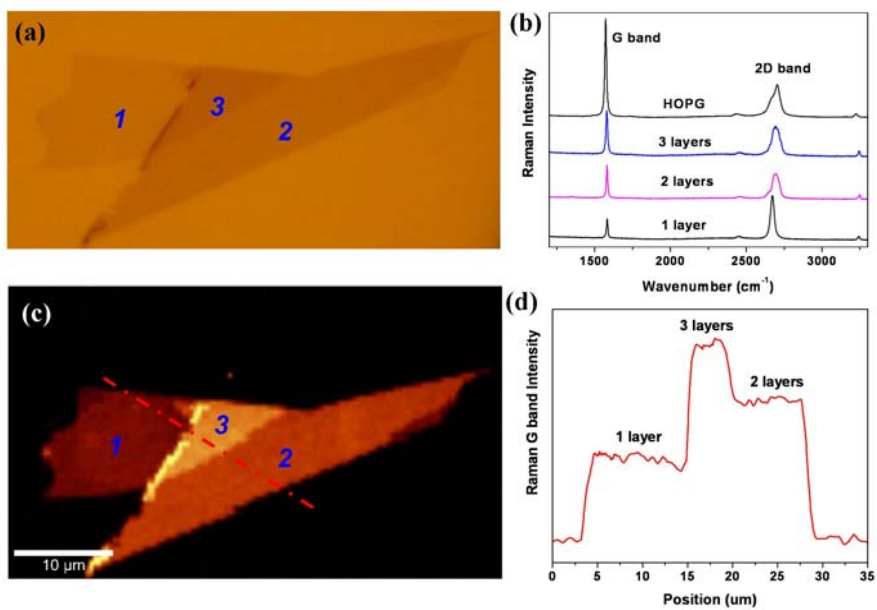

Figure 3. (a) Optical image of graphene with 1, 2, 3 and 4 layers. (b) Raman spectra of graphene with 1, 2, 3 layers as well as HOPG. (c) Raman image plotted by the intensity of G band. (d) The intensity cross section of Raman image, which corresponds to the dash lines.

Figure 3a shows the optical image of a graphene sample on the SiO$_2$/Si substrate. The graphene sheet contains one, two, and three layers, which are pre-determined by AFM. Fig 3b shows the Raman spectra of graphene with different thickness, as well as that of HOPG for comparison. Fig 3c gives the Raman imaging of G band intensity of the graphene sheets. The difference is very obvious for graphene with different thicknesses. The intensity of the G band along the dash line is shown in Fig. 3d. It can be clearly seen that the intensity of G band is almost linearly increase with the increase of graphene thickness. This can be used to determine layer numbers of multi-layer graphene. The advantage of Raman spectroscopy and imaging in determining number of layers of graphene is that it does not dependent on the substrate used. This is because Raman spectrum is the intrinsic characteristic of graphene.

In addition to the Raman spectroscopy/imaging, contrast spectroscopy can also be used to accurately identify the graphene layer numbers (up to 10 layers), which offers a cheaper alternative.[44] The contrast between graphene sheets and substrate is generated and quantified in contrast spectrum with the following calculation:

$$C(\lambda) = \frac{R_0(\lambda) - R(\lambda)}{R_0(\lambda)}, \quad (1)$$

By using the contrast method, the number of layers of unknown graphene sheet on 300 nm SiO$_2$/Si substrate can be determined precisely by using the following equation:[44]

$$C = 0.0046 + 0.0925N - 0.00255N^2 \quad (2)$$

where $N$ ($\leq 10$) is the number of layers of graphene sheet.

Although current research mainly focuses on the single and bilayer graphene, we believe that the few layers (less than 10 layers) graphene also have interesting properties as they still exhibit the two-dimensional properties.[1] By combining Raman spectroscopy and the contrast method, the graphene layer numbers can be unambiguously identified in any substrate, which helps future research and application of graphene.

## 4. Interference enhancement of graphene on SiO$_2$/Si substrate

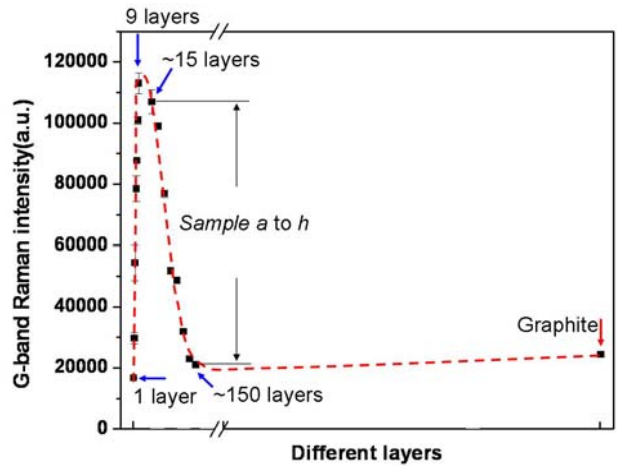

Figure 4. The G-band Raman intensity of graphene sheets as a function of number of layers. The red dashed curve is a guide to eye. Reproduced with permission from *Appl. Phys. Lett.* 2008, 92, 043121. Copyright 2008 American Institute of Physics. [40]

The very strong Raman signal from graphene (single and few layers) on SiO$_2$/Si substrate has been observed by many researchers and this makes the Raman studies of graphene available.[19, 37, 39] Normally, the Raman intensity is expected to be proportional to the thickness of graphene sample up to the laser penetration depth, which is around 50 nm (~150 layers) for 532 nm excitation. The penetration depth z of 532 nm laser in graphite is estimated by $\frac{I}{I_0} = e^{-\alpha z} = \frac{1}{e}$, where $\alpha = \frac{4\pi k}{\lambda}$, $\lambda$ =532nm, k=1.3 is the extinction coefficient of graphite.[45] However, we find that the G-band Raman signal of graphene sheets would decrease when the number of layers of graphene exceeds certain value (~10 layers). Figure 4 shows the G-band Raman intensity of graphene sheets on a 300 nm SiO$_2$/Si substrate as a function of number of layers.[40] The number of layers of graphene sheets are determined by Raman and contrast spectroscopy.[44, 46, 47] Besides the graphene sheets with one, two, three, five, six, eight, and nine layers, samples *a- h* are more than 10 layers, whose thickness are ~5 nm (sample *a*) and ~50 nm (sample *h*), respectively, as determined by AFM. Interestingly, it can be seen that the Raman intensity increases almost linearly until ~10 layers (~ 3 nm) and decreases for thicker graphite sheets. Besides, it is obvious that Raman signal of bulk graphite is weaker than that of bilayer graphene (BLG). These results deviate from the above understanding that laser penetration depths decide the thickness where highest Raman intensity can be obtained. Therefore, some mechanism must be responsible for this phenomenon.

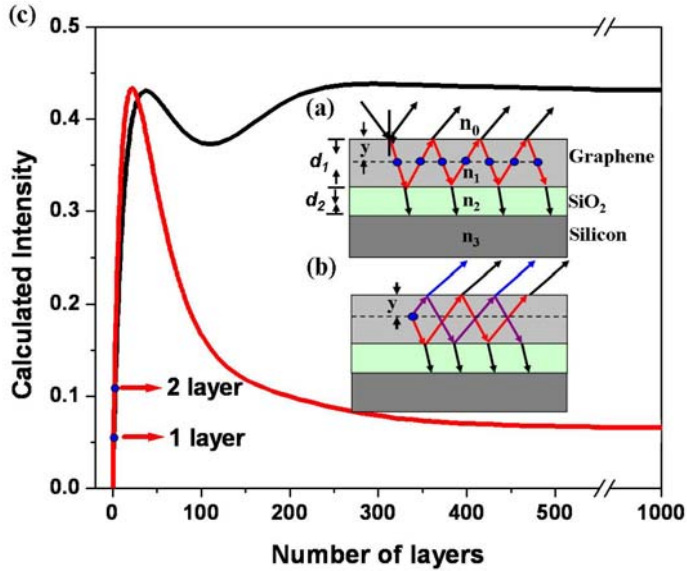

Figure 5. (a) Schematic laser reflection and transmission in certain depth *y* in graphene sheets deposited on SiO$_2$/Si substrate. (b) Multi-reflection of the scattering Raman light (from depth *y*) at the interface graphene/air and graphene/ SiO$_2$. (c).The calculation results of Raman intensity of G band as a function of number of layers with (red) and without (black) considering the multi-reflection of Raman scattering light in graphene. Reproduced with permission from *Appl. Phys. Lett.* 2008, 92, 043121. Copyright 2008 American Institute of Physics. [40]

Multi-reflection/interference effect on contrast has been well documented but the multiple reflection of Raman is not.[44, 46, 47] By considering the multilayer interference of incident light as well as the multi-reflection of Raman signals in graphene/graphite based on the Fresnel's equations, the above Raman results can be explained. Consider the incident light from air onto a graphene sheet, SiO$_2$/ Si trilayer system, as shown in Fig. 5(a). When a beam of light front at an interface, for example, the air/graphene or graphene/SiO$_2$ interface, a portion of the beam is reflected and the rest is transmitted, thus, an infinite number of optical paths are possible. $\tilde{n}_0$=1 is the refractive index of air. $\tilde{n}_1$=2.6-1.3i, $\tilde{n}_2$=1.46, $\tilde{n}_3$=4.15-0.044i, are refractive indices of graphite, SiO$_2$, and Si at 532 nm, respectively.[48] $d_1$ is the thickness of graphene which is estimated as $d_1=N\Delta d$, where $\Delta d =0.335$nm is the thickness of single layer graphene, and $N$ is the number of layers. $d_2$ is the thickness of SiO$_2$ and the Si substrate is considered as semi-infinite. The G-band Raman intensity of graphene sheets depends on the electric field distribution, which is a result of interference between all these transmitted optical paths in graphene/graphite sheets. The total amplitude of the electric field at certain depth $y$ in graphene/graphite sheets is viewed as a sum of the infinite transmitted laser, as schematically shown in Fig. 5(a). In addition, further consideration should be applied to the multi-reflection of scattering Raman light in graphene/graphite at the interface of graphene/air and graphene/SiO$_2$, which contribute to the detected Raman signal. Fig 5(b) schematically shows multi-reflection of the scattering light in certain depth $y$ in graphene sheets. Thus, the total Raman signal is a result of interference of transmitted laser followed by considering multi-reflection of Raman scattering light.[40]

Figure 5(c) shows the calculation results of Raman intensity of G band as a function of number of graphene/graphite layers. The black curve shows the calculation result without considering the multi-reflection of Raman light in graphene. It can be seen that the Raman scattering intensity is strongest at ~38 layers and the intensity of bulk graphite (~1000 layers) is much stronger than that of single and bilayer graphene, which is not agreed with experimental results. The red curve gives the calculated results after considering the multi-reflection of Raman light in graphene. In that case, the intensity is strongest at ~ 22 layers and the Raman signal of bulk graphite is weaker than that of BLG, which agrees very well with the experiment data. Thus, the consideration of the multi-reflection of scattering Raman light in graphene/graphite is necessary. Our calculation for Raman intensity of graphene can also be applied to other thin films and materials.

The idea of interference enhancement can also be applied to other thin film sample in order to enhance the Raman signal. A general selection of capping layer and substrate for enhancement can be achieved by choosing $n_1 \gg n_2 \ll n_3$, plus $n_2 d_2 = \frac{\lambda}{4}, \frac{3\lambda}{4}$, and so on.( $n_1$, $n_2$, $n_3$ are individually refractive indices of sample, capping layer and substrate). A total phase change of $2\pi$ ($\pi$ change at the interface of capping layer/ substrate plus $\pi$ due to the double thickness of capping layer) can be achieved by this kind of configuration. Therefore, all the transmitted light in graphene layers will have no phase change and in turn enhance Raman signal greatly.

**5. Raman studies of single layer graphene: the substrate effect**

In the following sections (5-7), we will individually discuss the effect of substrates, top layer deposition, annealing process, as well as folding on the atomic and electronic properties of graphene as probed by Raman spectroscopy and imaging.

Till now, most of the Raman studies were carried out on graphene sheets fabricated by micromechanical cleavage (MC) and transferred to Si substrate with appropriate thickness of $SiO_2$ capping layer (~300nm).[19, 39, 44, 47] However, different substrates might be used in different applications, such as the transparent substrates for optical application. Clear understanding of the interaction between graphene and substrate is important for potential applications and device fabrication of graphene. The systematical Raman study of single layer graphene (SLG) produced by MC on different substrates will be presented here.[43] Choosing SLG as the object is firstly due to the fact that it can be identified unambiguously by Raman spectroscopy from the characteristic 2D-band feature. Secondly, compared with graphene of a few layers, SLG is more sensitive to the interaction between graphene sheets and substrate. We also compared the Raman features of SLG on the above-mentioned substrates with those of epitaxial monolayer graphene (EMG) grown on SiC substrate, for which we believe there is a much stronger interaction between graphene and substrate.

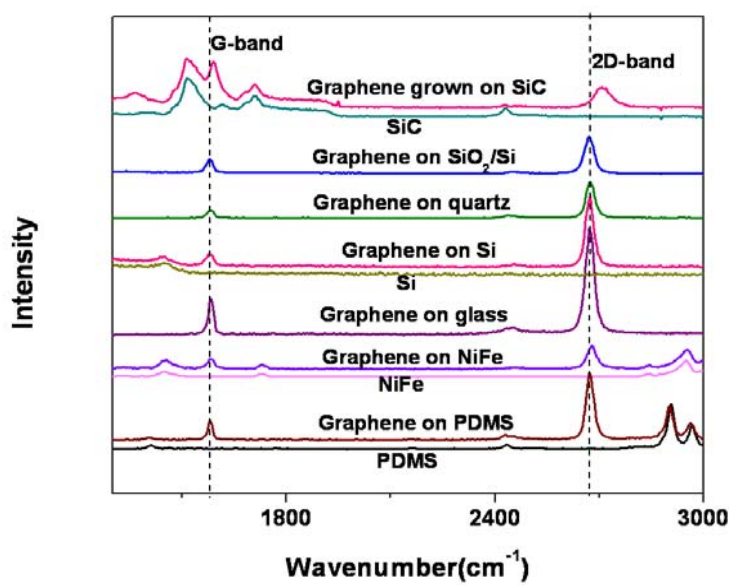

Figure 6 The Raman spectra of SLG on different substrates as well that of epitaxial monolayer graphene on SiC. Reproduced with permission from *J. Phys. Chem. C* 2008, *112*, 10637-10640. Copyright 2008 American Chemical Society.[43]

SLG is fabricated on different substrates, such as transparent substrates (glass and quartz), flexible substrates (polyethylene terephthalate (PET) and polydimethylsiloxane (PDMS)), conductive substrates (NiFe and highly doped Si), as well as the standard SiO$_2$/Si substrate. All these substrates are suitable for various applications. Figure 6 shows the Raman spectra of SLG on different substrate, from bottom to top, PDMS, NiFe, glass, Si, quartz and SiO$_2$(300nm)/Si substrate as well as the Raman spectrum of EMG grown on SiC substrate. The SLG graphene is identified by both the sharp 2D Raman band and the contrast method.[43] The G band and 2D band position and their full width at half maximum (FWHM) at different substrates are summarized in Table I. The result of graphene on GaAs substrate reported by Calizo et al [49] as well as graphene on SiO$_2$/Si substrate reported by Casiraghi et al. [50] is also included in Table I. One can see that G-band position (1581$\pm$1 cm$^{-1}$) and FWHM (15.5$\pm$1 cm$^{-1}$) are similar for graphene on SiO$_2$ (300nm)/Si, quartz, PDMS, Si, glass and NiFe substrates. The small difference in the G-band position on these substrates are within the range of fluctuation (1580-1588 cm$^{-1}$) by unintentional electron or hole doping effect reported by Casiraghi et al. [50] for more than 40 graphene samples on SiO$_2$/Si substrate. Therefore, our observation indicates that the interaction between micromechanically cleaved graphene and different substrates is not strong enough to affect the physical structure of graphene. These results are in line with Calizo et al [49, 51] who suggested that the weak substrate effect can be explained by the fact that G-band is made up of the long-wavelength optical phonons(TO and LO),[41] and the out of plane vibrations in graphene are not coupled to this in plane vibration.[52] On the other hand, for graphene grown on SiC substrate, it can be seen the intensity ratio of G and 2D band of EMG differs a lot from those of SLG made by MC. Moreover, significant blue shifts of G band (11 cm$^{-1}$) and 2D band (~ 34 cm$^{-1}$) of EMG are observed compared to those of graphene made by MC. This significant blueshift of Raman bands can be understood by the strain effect caused by the substrate. Between EMG and SiC substrate, there is an interfacial carbon layer, which has graphene like honeycomb lattice and covalently bonded to the SiC substrate.[53, 54] Such kind of bonding would change its lattice constant as well as the electronic properties. Therefore, the lattice mismatch between graphene lattice and interfacial carbon layer may cause a compressive stress on EMG, hence the shift of the G band Raman peak frequencies. The detail studies of epitaxial graphene on SiC substrate will be shown in section 8.[55]

Through the Raman studies of SLG produced by MC on different substrates, we could know the weak interaction between graphene sheets and the substrates play a negligible role in affecting the Raman features of graphene sheets. Only EMG grown on SiC substrate shows strong blue shift of G band which can be understood by the strain effect caused by the strong interaction between graphene and SiC substrate.

TABLE 1. The G-band and 2D-band position and their full width at half maximum (FWHM) for graphene/graphite on different substrates. Reproduced with permission from *J. Phys. Chem. C* 2008, *112*, 10637-10640. Copyright 2008 American Chemical Society. [43]

| Substrate | G band Position (cm$^{-1}$) | G band FWHM(cm$^{-1}$) | 2D band Position(cm$^{-1}$) | 2D band FWHM(cm$^{-1}$) |
|---|---|---|---|---|
| SiC | 1591.5 | 31.3 | 2710.5 | 59.0 |
| SiO$_2$/Si | 1580.8 | 14.2 | 2676.2 | 31.8 |
| SiO$_2$/Si [50] | 1580-1588 | 6-16 | | |
| Quartz | 1581.9 | 15.6 | 2674.6 | 29.0 |
| Si | 1580 | 16 | 2672 | 28.3 |
| PDMS | 1581.6 | 15.6 | 2673.6 | 27 |
| Glass | 1582.5 | 16.8 | 2672.8 | 30.8 |
| NiFe | 1582.5 | 14.9 | 2678.6 | 31.4 |
| GaAs [49] | 1580 | 15 | | |
| Graphite | 1580.8 | 16.0 | 2D$_1$: 2675.4<br>2D$_2$: 2720.8 | 41.4<br>35.6 |

**6. Process induced defects and strain on graphene**

Besides global back gates, top local gates are proposed for more complex graphene devices, such as pn junction,[56] Veselago lens [57] and Klein tunneling.[58] The top gate oxides that have been used so far include $H_fO_2$, $Al_2O_3$ and $SiO_2$. Although efforts have been made to deposit the gate oxides without damaging the graphene or changing its electrical properties,[56-58] the gate oxides should influence the graphene sheets in at least three ways: doping, defects, and various mechanical deformations. We will present the systematical study of graphene sheets subjected to defects and mechanical deformations induced by insulating capping layers using Raman spectroscopy and imaging.[59]

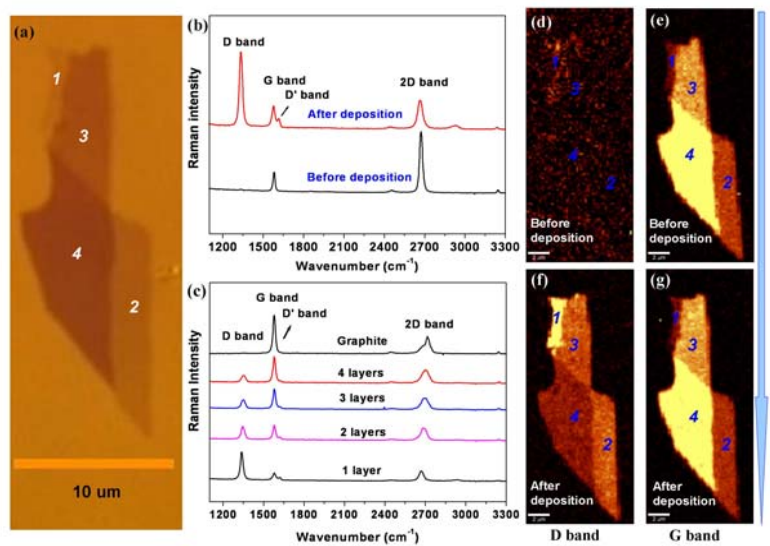

Figure 7. (a) Optical image of graphene sheets on $SiO_2$/Si substrate. (b) Raman spectra of SLG before and after the 5 nm $SiO_2$ deposition. (c) Raman spectra of graphene with different thicknesses as well as that of bulk graphite after $SiO_2$ deposition. Raman images of graphene sheets before $SiO_2$ deposition generated from the intensity of the D band (d) and G band (e). Raman images of graphene sheets after 5 nm $SiO_2$ deposition using the intensity of D band (f), and G band (g). Reproduced with permission from *ACS Nano* 2008, 2, 1033-1039. Copyright 2008 American Chemical Society. [59]

Figure 7(a) shows the optical image of a graphene sample on the $SiO_2$/Si substrate. The graphene sheet contains one, two, three and four layers as determined by Raman and contrast methods.[44] A 5 nm $SiO_2$ top layer was then deposited on graphene by Pulse laser deposition (PLD) with a 248 nm KrF pulsed laser. The laser power used was very weak (~200 mJ and repetition rate of 10Hz) to achieve the slow and smooth deposition (1Å/min). The Raman spectra of graphene before and after $SiO_2$ deposition were shown in Figure 7(b). A clear difference is that two defect-induced Raman bands, D (1350 $cm^{-1}$) and D′ (1620 $cm^{-1}$) band,[42, 60] were observed after deposition. The observation of D and D′ bands indicate that defects were introduced into graphene after the 5 nm $SiO_2$ top layer deposition. This may be caused by the damage on the sample during deposition, or by the interaction between $SiO_2$ and graphene which may produce vacancy, dislocation and/or dangling bonds. Annealing is carried out to eliminate the defects, which will be discussed in later. Figure 7(c) shows the Raman spectra of graphene sheet with one to four layers as well as that of bulk graphite after $SiO_2$ deposition. The Raman spectra were taken under same conditions. The D band intensity decreases with the increase of graphene thickness and is invisible for bulk graphite, demonstrating that defects are more easily introduced into thinner graphene sheets.[37] Figure 7(d) and 7(f) show Raman images generated from the intensity of D band before and after deposition respectively. Before deposition, there is no D band hence the Raman image is dark. After deposition, the thinner graphene (single layer graphene) shows the stronger D band, which is consistent with the discussion above. Figure 7(e) and 7(g) show the images generated from the intensity of the corresponding G band, and they do not show noticeable difference.

We have also deposited different materials as capping layers ($SiO_2$, $HfO_2$, Si and polymethyl methacrylate (PMMA)) with different methods, such as electron beam evaporation, PLD, RF sputtering, and spin coating. Our results show that the deposition methods have a significantly effect on the defects, with spin coating introducing the least amount of defects (no D band), followed by electron beam evaporation (weak D band). PLD and RF sputtering introduce the most defects (strong D band).

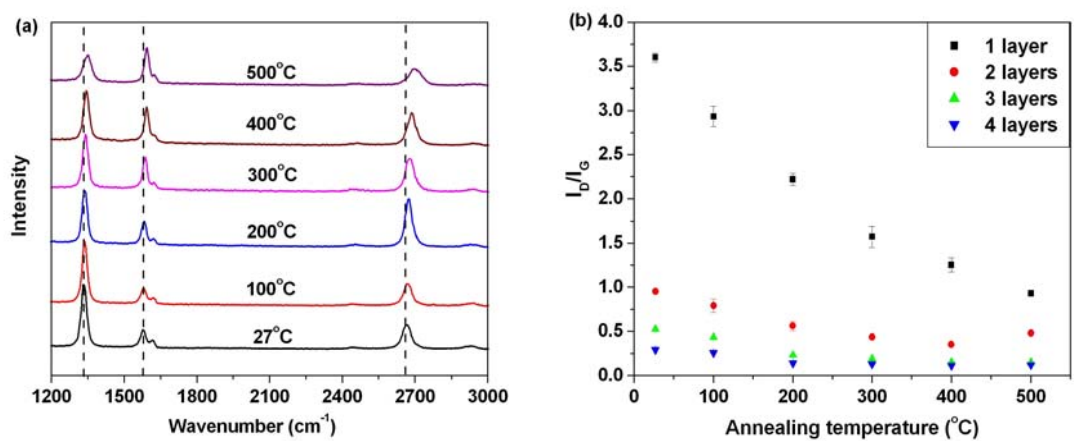

Figure 8. (a) Raman spectra of single layer graphene coated by 5 nm $SiO_2$ and annealed at different temperature. (b) The intensity ratio of D band and G band ($I_D/I_G$) of graphene sheets with one to four layers (coated with $SiO_2$) after annealing at different temperature. The $I_D/I_G$ (defects) decreased significantly upon annealing. Reproduced with permission from *ACS Nano* 2008, 2, 1033-1039. Copyright 2008 American Chemical Society. [59]

The Raman spectra of single layer graphene after $SiO_2$ deposition and annealed in air ambient at different temperatures are shown in Figure 8(a). We have also carried out vacuum annealing and similar results were observed. An obvious observation is that the intensity of D band decreases upon annealing. This is clearly demonstrated in Figure 8(b), which shows the intensity ratio between the D band and G band ($I_D/I_G$) that is often used to estimate the amount of defects in carbon materials. For one to four-layer graphene sheets, this ratio decreases with increase in annealing temperature. This can be understood as due to the recovery of damaged graphene at high temperature. Another important observation can be seen in Fig. 8(a) is that all the Raman bands shifted to higher frequency with increase in annealing temperature. Fig. 9(a) shows the blueshift of G band of graphene sheets as the annealing temperature increases. The G band blue shifted ~15 cm$^{-1}$, while the D band blue shifted ~13 cm$^{-1}$ and 2D band ~29 cm$^{-1}$ after annealing at 500 $^o$C. We attribute this significant blueshift of Raman bands to the strong compressive stress on graphene. The $SiO_2$ becomes denser upon anneal so it exerted a strong compressive stress on the graphene. For comparison, the Raman bands of bulk graphite did not shift after deposition and annealing, which supported the above explanation, as bulk graphite is too thick and it is not easily compressed by SiO$_2$. Recently, Yan et al.[22] and Pisana et al.[23] found that the frequency of the G and 2D Raman bands can also be adjusted by charge doping through electron-phonon coupling change. Besides the G band blueshift, a bandwidth narrowing of ~10 cm$^{-1}$ was also observed in the case of charge doping. However, in our results, only a small fluctuation (±1 cm$^{-1}$) of G band FWHM (full width at half maximum) was observed after annealing at different temperature, which indicates that the effect of charge doping can be ignored. In addition, it is shown that the dependence of the 2D band blueshift on doping is very weak and only ~10-30% compared to that of G band.[22, 25] The 25 cm$^{-1}$ blueshift of the 2D band is too large to be achieved by charge doping alone. Therefore, the observed shifts of G (~15 cm$^{-1}$) and 2D (~25 cm$^{-1}$) band in the above experiment were mainly caused by stress.

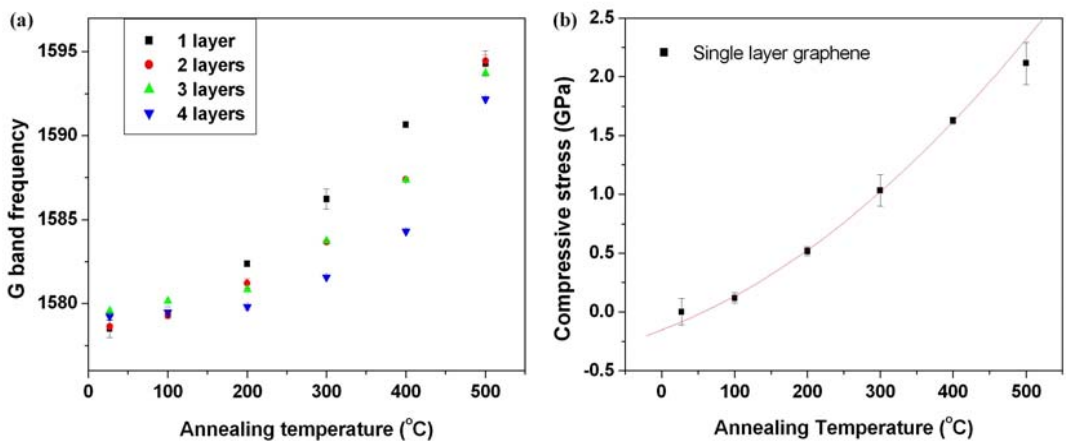

Figure 9. The Raman peak frequency of G band of graphene sheets with one to four layers (coated with SiO$_2$) after annealing at differenet temperature. (b) Magnitude of compressive stress on single layer graphene controlled by annealing temperature. The red line is a curve fit to the experimental data. Reproduced with permission from *ACS Nano* 2008, 2, 1033-1039. Copyright 2008 American Chemical Society. [59]

The compressive stress on graphene in the above experiment is due to the denser of SiO$_2$ upon annealing. This origin of the compressive stress is very similar to the biaxial stress due to the lattice mismatch at the sample/substrate interface in a normal thin film. Therefore, the stress on graphene should be biaxial. The biaxial compressive stress on graphene can be estimated from the shift of Raman $E_{2g}$ phonon. From a biaxial stress model, the G band stress coefficient of graphene is estimated to be 7.47 cm$^{-1}$/GPa.[59] The stress on single layer graphene with annealing temperature can then be estimated and it is shown in Figure 9(b). The compressive stress on graphene was as high as ~2.1 GPa after depositing SiO$_2$ and annealing at 500 °C, and the stress on single layer graphene in our experiment can be fitted by the following formula:

$$\sigma = -0.155 + 2.36 \times 10^{-3} T + 5.17 \times 10^{-6} T^2 \qquad (3)$$

where σ is the compressive stress in GPa and T is temperature in °C. The appearance of such large stress is mainly because graphene sheets are very thin (0.325 nm in thickness for single layer graphene),[61] so that they can be easily compressed or expanded. We have also introduced tensile stress onto graphene by depositing a thin cover layer of silicon. The G band of graphene red shifted by ~5 cm$^{-1}$ after silicon deposition, which corresponds to a tensile stress of ~0.67 GPa on graphene sheet. We suggest that tensile stress can be also achieved by depositing other materials with larger lattice constant than graphene. In combination with annealing, both compressive and tensile stress can be introduced and modified in graphene in a controllable manner. The stressed graphene may have very important applications as the properties of graphene (optical and electronic properties) can be adjusted by stress, where stress studies in CNTs have already set good example.[62-64] Stress engineering using SiGe alloy has already been used in the IC fabrication to improve the device performance.

## 7. Raman spectroscopy of folded graphene

The peculiar properties of graphene are closely related to its unique electronic band structure. For single layer graphene (SLG), it is known that the low energy dispersion is linear,[65] which make charge carriers in SLG behave like massless Dirac fermions. Interestingly, recent studies such as magnetotransport,[31] far infrared magneto-transmission [66] investigations on multi-layer epitaxial graphene (EG) with stacking disorder [67, 68] still revealed the two dimensional Dirac-like (SLG-like) character of electronic states. Furthermore, theoretical calculations of electronic structure of BLG with a twist of second layer were carried out. The results show that the low energy dispersion of twisted two-layer graphene is linear, as in SLG, but the Fermi velocity is significantly smaller.[69]

Further investigation on the electronic structure of BLG that deviates from the AB stacking is necessary to obtain fundamental understanding of the relation between stacking order and electronic properties of bi- or multi- layer graphene.

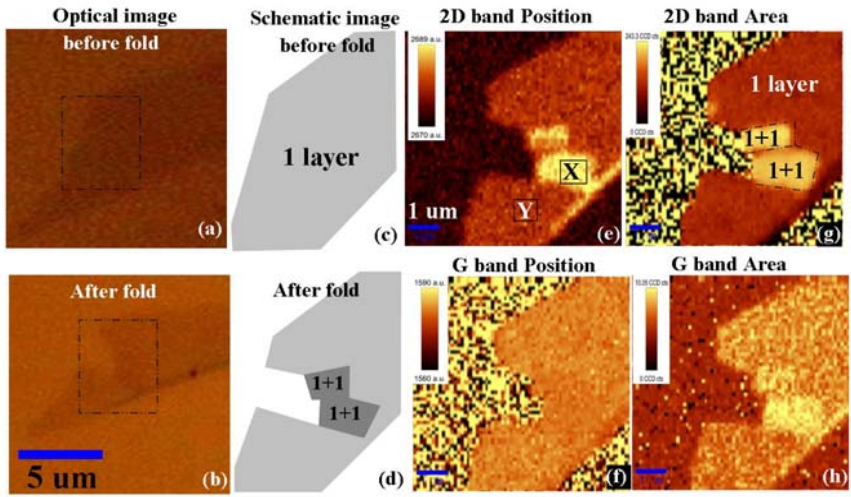

Figure 10(a) and (b) show the optical image of SLG before and after fold. (c) and (d) give the schematic image of SLG before and after fold. (e) and (f) respective shows Raman image obtaining from the 2D and G band positions. (g) and (h) individually show Raman image by extracting the area of 2D and G band. Reproduced with permission from *Phys. Rev. B* 2008, 77, 235403. Copyright 2008 American Physical Society. [70]

We have successfully fabricated the 1+1 folded graphene with different stacking order and studied their electronic properties using Raman spectroscopy.[70] The 1+1 layer samples are prepared by simply gently flushing de-ionized water across the surface of the substrate containing the target graphene sheet. The 1+1 layer folded graphenes can be observed after this process. Fig. 10(a) and (b) show the optical images of SLG before and after folding. The black dashed rectangle indicates the location where folding happens. As the optical images are not very clear, Fig. 10(c) and (d) show the schematic drawing of the sample before and after folding. Fig. 10(e) and (f) respectively show Raman images of 2D and G band position (frequency) after folding. As shown in Fig. 10(e), the 2D band frequency of folded graphene is much higher than that of SLG. On the other hand, in Fig. 10(f), even contrast can be seen which means almost no change of G band frequency before and after folding. Fig. 10(g) shows the Raman image obtained from the 2D peak area (integrated intensity) after folding. The intensity of folded graphene is much higher than that of SLG. It is observed in the previous work [39] that the area of the 2D band is almost identical for 1 to 4 layers graphene. Hence, the much stronger 2D band of folded graphene reveals its structure and properties are different from that of BLG. Fig. 10(h) gives the Raman image extracted from the G band area after folding. The G band intensity of folded graphene is nearly double of SLG since it contains two layers of carbon atoms.

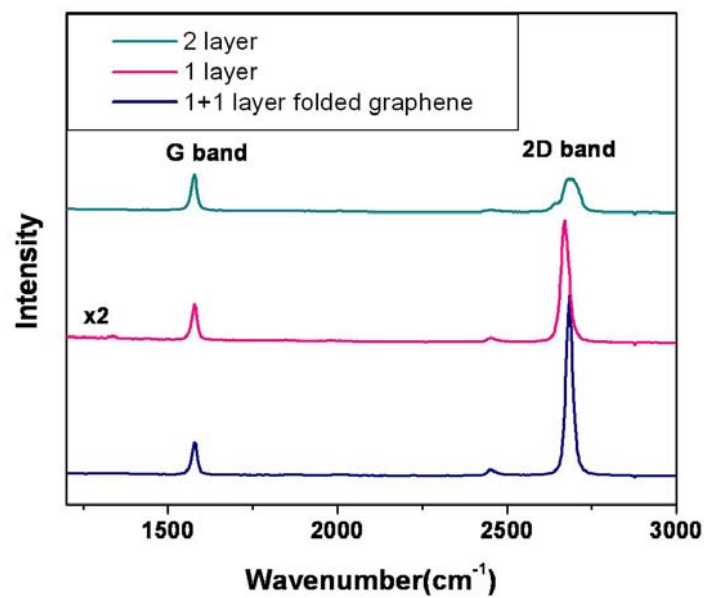

Figure 11 the Raman spectrum of BLG, SLG and 1+1 layer folded graphene. The spectra are normalized to have similar G band intensity. Reproduced with permission from *Phys. Rev. B* 2008, 77, 235403. Copyright 2008 American Physical Society. [70]

Figure 11 gives typical Raman spectra of the 1+1 folded graphene (taking from area **X** in Fig. 10(e)), SLG (taking from area **Y** in Fig. 10(e)) and BLG for comparison. The spectra are normalized to have similar G band intensity. From this figure, it can be obviously seen that the Raman spectrum of 1+1 folded graphene is different from that of BLG. The 2D band of BLG is much broader and can be fitted as 4 peaks, which originates from splitting of valence and conduction bands.[19] However, for 1+1 folded graphene, only a single sharp peak exists which is similar to that of SLG. Thus, the electronic structure of folded graphene should be similar to that of SLG, i.e. there is no splitting of energy bands. The similar 2D band of folded graphene and SLG calls for attention when using Raman to identify SLG. Although the Raman features of folded graphene is quite similar as that of SLG, there are differences needed to be noticed: A strong blue shift (~12 cm$^{-1}$) of the 2D band of folded graphene (2686 cm$^{-1}$) compared to SLG (2674 cm$^{-1}$) can be clearly seen, as indicated in the Raman images (Fig. 10(e)). This blue shift is associated with the SLG-like band structure of 1+1 folded graphene but with smaller Fermi velocity, which will be discussed in detail latter. The folded graphene has higher 2D to G band intensity ratio than that of SLG, partially due to the different resonance conditions of folded graphene and SLG. We have studied a total of six 1+1 layer folded graphene samples. Blue shift of the 2D band compared to SLG is observed for every sample, but the amount of blue shift differs from 4 to 14 cm$^{-1}$.

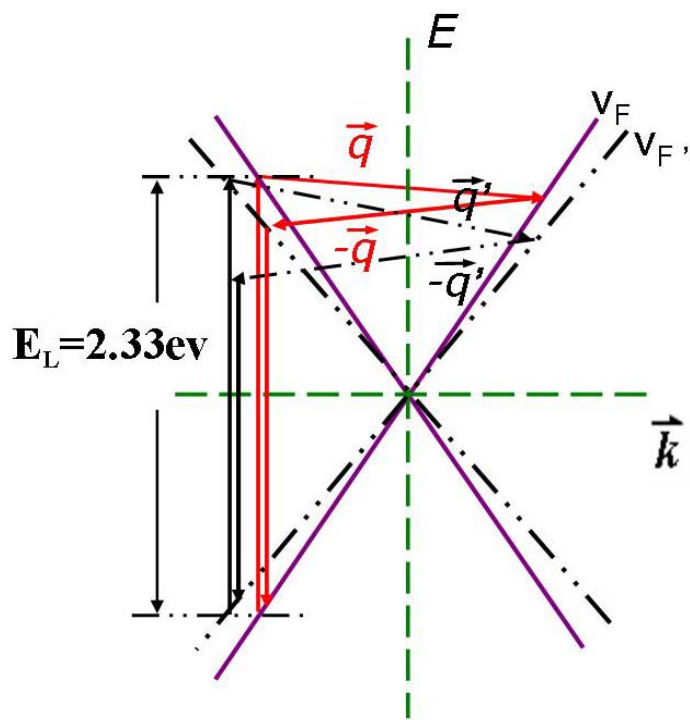

Figure 12 schematically shows the electronic structure of 1+1 layer folded graphene (black dash lines) as well as SLG (purple solid lines). The arrows indicate the double resonance process in folded graphene (black dash lines) and SLG (red solid lines). Reproduced with permission from *Phys. Rev. B* 2008, 77, 235403. Copyright 2008 American Physical Society. [70]

Lopes dos Santos *et al.* [69] calculated the electronic structure of twisted BLG and their results show that it has SLG-like linear dispersion with slowing down of Fermi velocity. The 1+1 folded graphene can be viewed as twisted BLG with unknown twisted angle. Therefore, the energy band structure of our folded graphene is estimated under the model of twisted BLG. Fig. 12 schematically shows the electronic structure of 1+1 folded graphene as well as SLG. The slowing down of Fermi velocity ($v_F$') of folded graphene corresponds to the smaller slope of energy dispersion near Dirac point. The second-order double resonance Stokes process for folded graphene and SLG are schematically shown in Fig. 12. For folded graphene, phonon with larger wave vector ($q$') is needed to inelastically scatter the electron compared to that of SLG ($q$). Because of the almost linear dispersion of optical phonon branch around K point ($\frac{d\omega}{dq}$)[19] which contributes to the observed 2D band frequency, phonon with higher frequency ($\omega$') is obtained in double resonance process for folded graphene. This is the reason of blue shift of the 2D band of folded graphene. By using this model, the sharpness and blueshift of 2D band of multilayer EG on carbon terminated SiC can be also easily understood,[71] which are caused by the SLG-like linear dispersion of energy band of multilayer EG but with ~10% slowing down of Fermi velocity [66, 72] than that of SLG.

Moreover, the Fermi velocity of folded graphene can be estimated. In the double resonance process,[42] it is known that

$$\hbar v_F q \approx E_L - \hbar(\omega/2) \tag{4}$$

Here, $E_L$ is the incident photon energy, $v_F$ is the Fermi velocity, $\omega$ is the frequency of two phonon 2D band, $\hbar(\omega/2)$ is the phonon energy.

With simple mathematics, the relation of Fermi velocity change and 2D frequency shift can be achieved as: [70]

$$\frac{dv_F}{v_F} = -\frac{\hbar v_F}{[E_L - \hbar(\omega/2)] \cdot \frac{d\omega}{dq}} d\omega \tag{5}$$

Here, $\hbar v_F = 6.5 eV \text{ Å}$,[42] $E_L$=2.33eV, $\hbar(\omega/2) = 0.166 eV$ by taking $\omega$≈2670 cm$^{-1}$. $\frac{d\omega}{dq} = 645 cm^{-1} \text{ Å} = 0.08 eV \text{ Å}$ is the phonon dispersion around K point.[19]

Therefore,

$$\frac{dv_F}{v_F} = -\frac{37.5 \cdot d\omega}{eV} \quad or \quad -\frac{0.00467 \cdot d\omega}{cm^{-1}} \tag{6}$$

where $\frac{dv_F}{v_F}$ is the Fermi velocity change in percentage and $d\omega$ is the frequency change of 2D band after fold in eV or cm$^{-1}$.

For the 1+1 folded graphene sample, the blue shift of 2D band is ~12 cm$^{-1}$, thus, the slowing down of Fermi velocity is about 5.6%. The smaller blue shift (~4 cm$^{-1}$) corresponds to the smaller slowing down of Fermi velocity (~2%). This difference may be due to the different twist angle between the first and second graphene layer for different samples. Different twist angle between the two layers will result in different amount of slowing down of Fermi velocity,[69] hence the different blueshift of 2D band. Another reason is the separation of the two layers, which will also affect the electronic properties of folded graphene.

## 8. Raman spectroscopy of epitaxial graphene

Most of the Raman studies above were carried out on micromechanical cleavage graphene (MCG).[19, 37, 44] In this section, we present the Raman studies of epitaxial graphene (EG) grown on SiC substrate.[55] EG grown on SiC is suitable for large area fabrication is more compatible with current Si processing techniques for future applications. Nevertheless, the EG may interact with the SiC substrate which could modify its optical and electronic properties. A bandgap of ~0.26 eV was observed by angle-resolved photoemission spectroscopy (ARPES) on EG grown on SiC substrate, which attributed to the interaction of graphene with the substrate.[73] Some theoretical [54, 74] and experimental studies on EG, e.g. X-ray diffraction (XRD)[54, 75] and scanning tunneling microscopy (STM)[76, 77], have also been carried out. However the effect of SiC substrate on EG is still not well understood. The formation of graphene on SiC substrate can be described as follows: the SiC surface first reconstructs to a ($\sqrt{3} \times \sqrt{3}$)R30° (R3) structure, then to a ($6\sqrt{3} \times 6\sqrt{3}$)R30° (6R3) structure, referred as carbon nanomesh here, after higher temperature annealing, the graphene forms on carbon nanomesh. However, it is still under debate as to how the graphene bonds/connects to the 6R3/carbon nanomesh structure.

The EG samples in this experiment are prepared by the following process: annealing a chemically etched (10% HF solution) n-type Si-terminated 6H-SiC (0001) sample (CREE Research Inc.) at 850 $^0$C under a silicon flux for 2 min in ultrahigh vacuum (UHV) resulting in a Si-rich 3x3-reconstructed surface, and subsequently annealing the sample several times at 1300 $^0$C in the absence of the silicon flux to form EG.[31, 78] The structure of EG is confirmed by in-situ Low-Energy-Electron- Diffraction (LEED), STM, and photoemission spectroscopy (PES).[76] Since the characteristic STM images of carbon nanomesh and single layer graphene are quite different, the appearance of single layer EG can be determined by monitoring the phase evolution from carbon nanomesh to graphene by in-situ STM during the thermal annealing of SiC. [76, 79] However, the STM images for single layer and bilayer EG on SiC are very similar. It is very hard to determine the layer thickness using this method. Instead, layer thickness for bilayer or thicker graphene sample is measured by monitoring the attenuation of the bulk SiC related Si 2p PES signal (photon beam energy is 500 eV) with normal emission condition.[76]

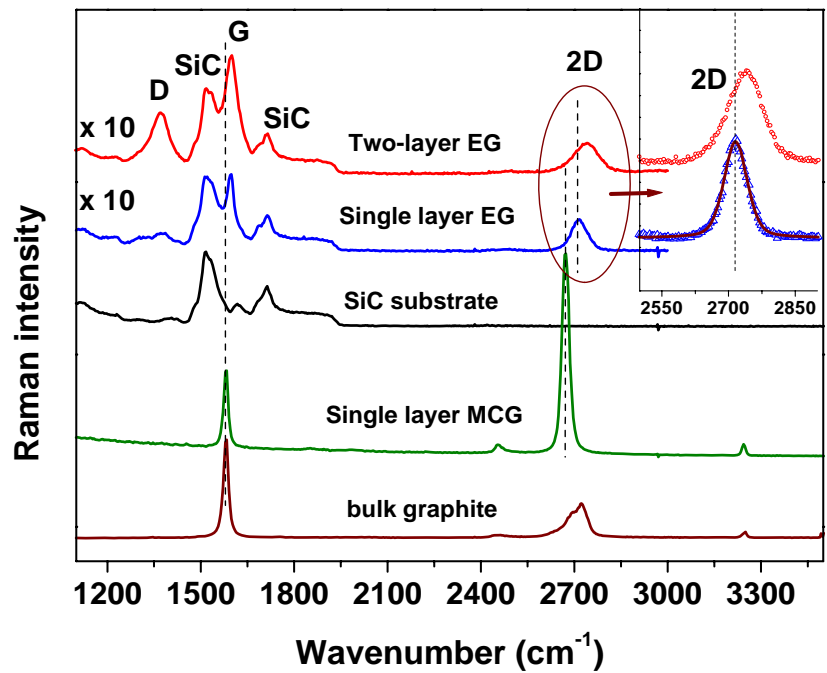

Figure 13. Raman spectra of single and two-layer EG grown on SiC, SiC substrate, MCG, and bulk

graphite as indicated. The inset is an enlarged part of the 2D band region of single and two-layer EG. The hollow symbols are experimental data and the solid line is the fitted curve. Reproduced with permission from *Phys. Rev. B* 2008, 77, 115416. Copyright 2008 American Physical Society. [55]

Figure 13 shows the Raman spectra of single and two-layer EG (grown on Si terminated SiC), MCG, bulk graphite, and SiC substrate.[55] The 6H-SiC has several overtone peaks in the range of 1000 to 2000 cm$^{-1}$. The peak near ~1520 cm$^{-1}$ is the overtone of the TO(X) phonon at 761 cm$^{-1}$. The peak near ~1713 cm$^{-1}$ is a combination of optical phonons with wave vectors near the M point at the zone edges.[80, 81] The weak SiC peak at ~1620 cm$^{-1}$ is not observable in our EG samples. The Raman spectrum of single layer EG has five peaks, located at 1368, 1520, 1597, 1713, and 2715 cm$^{-1}$, of which the peaks at 1520 and 1713 cm$^{-1}$ are from the SiC substrate. The 1368 cm$^{-1}$ peak is the so-called defect-induced D band; the 1597 cm$^{-1}$ peak is the in-plane vibrational G band; and the 2715 cm$^{-1}$ peak is the two-phonon 2D band.[60] The Raman signal of single layer MCG is much stronger (~10 times) than that of EG on SiC substrate. This can be explained by the interference enhancement of Raman signal of graphene on 300 nm SiO$_2$/Si substrate.[40] Compared with MCG and graphite, the Raman spectrum of EG shows the defect-induced D band, indicating that it contains defects, which may result from the surface dislocations, the corrugation, the interaction of graphene with substrate, or vacancies. The 2D band of single layer EG is broader than that of MCG, which is 60 cm$^{-1}$ compared to 30 cm$^{-1}$,[19] which can be explained by the poorer crystallinity of EG. However, compared to two- layer EG, the 2D band of single layer EG is still much narrower (60 cm$^{-1}$ compared to 95 cm$^{-1}$) and has a lower frequency (2715 cm$^{-1}$ compared to 2736 cm$^{-1}$), which are characteristics of single layer graphene. The Raman results confirm again that the EG on SiC in Fig. 13 is single and two layers, agree with the STM and PES identification. Another important observation is that the G (1597 cm$^{-1}$) and 2D (2715 cm$^{-1}$) bands of single layer EG shift significantly towards higher frequency from those of G (1580 cm$^{-1}$) and 2D (2673 cm$^{-1}$) of single layer MCG. Although the G band of single and few layer MCG may fluctuate ($\pm 3$ cm$^{-1}$) around the frequency of bulk graphite G band (1580 cm$^{-1}$), while the 2D bands of MCG may locate between 2660 and 2680 cm$^{-1}$,[39] the significant shifts of G band (17 cm$^{-1}$) and 2D band (42 cm$^{-1}$) of EG should be due to other mechanisms. The possibility that local electron/hole doping[22, 23, 25] in EG causes this Raman blueshift is not high, because the 42 cm$^{-1}$ 2D-band shift is too large to be achieved by electron/hole doping. Here, we attribute it to the interaction of SiC substrate with EG, most probably the strain effect, whereby the strain changes the lattice constant of graphene, hence the Raman peak frequencies.[55]

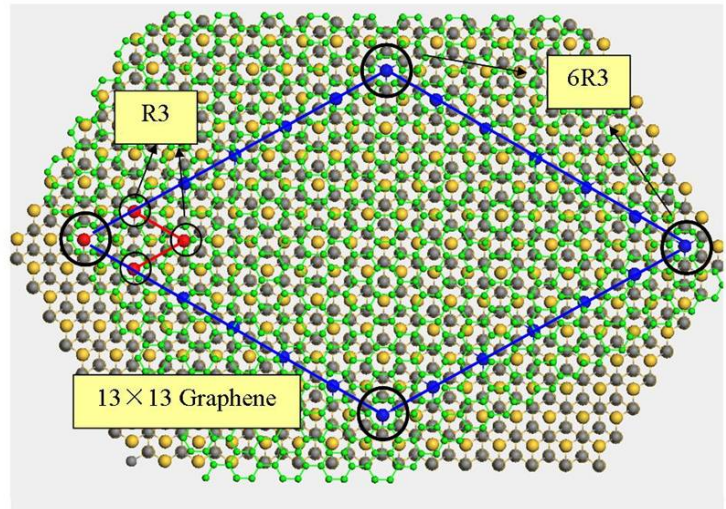

Figure 14. Schematic image of a graphene layer on SiC (0001) surface. The green, yellow and gray spheres represent C in graphene, Si in SiC and C in SiC, respectively. The SiC surface was after 6R3 reconstruction and a 13×13 graphene lattice lies on above it. The small black circles represent the R3 lattice, while the large black circles represent the 6R3 lattice. Reproduced with permission from *Phys. Rev. B* 2008, 77, 115416. Copyright 2008 American Physical Society.[55]

To illustrate the origin of the strain, Figure 14 shows the schematic image of a graphene layer on SiC (0001) 6R3 reconstructed surface. The large black circles represent the 6R3 lattice. The bulk lattice constant we used for SiC is 3.073 Å,[82] while that for graphene is 2.456 Å.[83] It is obvious that the 13 × 13 graphene (31.923 Å) matches the 6R3 lattice (31.935 Å) quite well.[53] On the other hand, the 2 × 2 graphene (4.9 Å) does not match the R3 structure (5.34 Å) (small black circles). Previous STM results showed that the 6R3 surface/carbon nanomesh did not always retain its "6 × 6" periodicity. The pore size of honeycombs in STM can be changed from 20 Å to 30 Å, depend on the annealing temperature.[76] Hence, this surface can be described as a dynamic superstructure formed by the self-organization of surface carbon atoms at high temperatures. As a result, the mismatch between graphene 13 × 13 lattice (~32 Å) and 6R3 surface/carbon nanomesh (20 to 30 Å) will cause the compressive strain on EG. Recently, Rohrl et al.[84] also observed the blueshift of Raman bands of EG and they attributed it to the compressive strain caused by the different thermal expansion coefficients between EG and SiC. This might be another origin of compressive strain. However, the reason of lattice mismatch (32 Å compared to 20-30 Å) between graphene and carbon nanomesh cannot be excluded.[84]

Graphene has a very thin (2D structure) and its stress induced by the lattice mismatch with the SiC substrate can be considered as biaxial. The biaxial compressive stress on EG can also be estimated from the shift of Raman $E_{2g}$ phonon (G band) with the stress coefficient about 7.47 cm$^{-1}$/GPa.[55] Hence, a biaxial stress of 2.27 GPa on EG is obtained from the 17 cm$^{-1}$ shift of G band frequency of EG compared to that of bulk graphite or MCG. The strong compressive stress may affect both the physical and electronic properties of graphene analogy to what happened in CNTs.[63, 64] Raman spectra of few layer epitaxial graphene on SiC substrate are also reported by Faugeras et al. [71] and no blueshift of the G band was observed. This may be because the expitaxial graphene they used is too thick (five to ten layers) and the effect of substrate on graphene is too weak to be observed as a consequence.

## 9. Conclusion

A review of the Raman spectroscopy and imaging of graphene is presented. The Raman and contrast spectroscopy is used as an unambiguous method to determine the number of layers of graphene. The very strong Raman signal of single layer graphene which is comparable to that of graphite is explained by an interference enhancement model. The effect of substrates on graphene has also been investigated. The results suggest that the weak interaction between graphene sheets and the substrates play a negligible role in affecting the Raman features of graphene sheets made by micromechanical cleavage. We also present the first experimental study of process-induced defects and stress in graphene using Raman spectroscopy and imaging. While defects lead to the observation of defect-related Raman bands, stress causes shift in phonon frequency. A compressive stress (as high as 2.1 GPa) is induced in graphene by depositing a 5 nm $SiO_2$ followed by annealing. Following, Raman spectroscopy is used to investigate the electronic structure of 1+1 folded graphene that deviated from AB stacking. The 1+1 folded graphene has a similar 2D band (sharp and symmetry) like SLG but has a buleshift (4 to 14 cm$^{-1}$), which is due to the SLG-like electronic structure but with reduction of Fermi velocity. The similar 2D band of folded graphene and SLG calls for attention when using Raman to identify SLG. Finally, the Raman spectroscopy of epitaxial graphene grown on SiC substrate is carried out. All the Raman peaks of epitaxial graphene have been assigned and compared with those of micromechanical cleavaged graphene and bulk graphite. The results show that graphene grown on SiC is compressive stressed.